\documentclass[12pt]{article}
\usepackage[utf8]{inputenc}
\usepackage[T1]{fontenc}
\usepackage[linktocpage=true,plainpages=false]{hyperref}
\usepackage{color}
\usepackage{amsmath,amssymb}
\usepackage{cite}
\usepackage{array}
\usepackage[labelsep=period]{caption}
\usepackage[title]{appendix}

\definecolor{linkcolor}{rgb}{0.6,0,0}
\definecolor{citecolor}{rgb}{0,0.6,0}
\definecolor{urlcolor}{rgb}{0,0,0.9}
\hypersetup{colorlinks, linkcolor={linkcolor},citecolor={citecolor}, urlcolor={urlcolor}}

\newcommand{\ls}{\left(}
\newcommand{\rs}{\right)}
\newcommand{\al}{\alpha}
\newcommand{\be}{\beta}
\newcommand{\ff}{\varphi}
\newcommand{\te}{\theta}

\begin{document}
\title{{Global embeddings of BTZ and Schwarzschild-AdS type black holes in a flat space}}
\author{A.~A.~Sheykin\thanks{E-mail: a.sheykin@spbu.ru}, D.~P.~Solovyev\thanks{E-mail: dimsol42@gmail.com}, S.~A.~Paston\thanks{E-mail: pastonsergey@gmail.com}
\\
{\it Saint Petersburg State University, Saint Petersburg, Russia}
}
\date{\vskip 15mm}
\maketitle

\begin{abstract}
	We study the problem of construction of global isometric embedding for spherically symmetric black holes with negative cosmological constant in various dimensions. Firstly, we show that there is no such embedding for 4D RN-AdS black hole in 6D flat ambient space, completing the classification which we started earlier. 	Then we construct an explicit embedding of 	{non-spinning} BTZ black hole in 6D flat ambient space. Using this embedding as an anzats, we then construct a global explicit embedding of $(d)$-dimensional Schwarzschild-AdS black hole in a flat $(d+3)$-dimensional ambient space.
\end{abstract}

\newpage

\section{Introduction}
Black holes are one of the most favorite objects of study in the general relativity. They are simple enough to be described only by a few parameters (for a historical survey of no-hair theorem see, e.g. \cite{nohair}) but nevertheless have a very rich structure and many interesting properties. Shortly after the appearance of the renowned Schwarzschild paper \cite{Schwarzschild}, his result was generalized  by Kottler \cite{Kottler}, who introduced a cosmological constant, and Reissner \& Nordstr$\ddot{\text{o}}$m \cite{Reissner}, who considered a charged point source.

The $n$-dimensional generalization of the black hole metric was found by Tangherlini \cite{Tangherlini}, who also mentioned that the absence of propagating degrees of freedom in (2+1)D gravitation with point masses is well-known\footnote{
	It is worth noting that this can be shown by a  simple argument based on the dimensionality of the Newtonian gravitational constant. Indeed, if we assume the satisfaction of Gauss theorem for Newtonian gravitational field in $n$ spatial  dimensions, then $F\propto G_n m M/r^{n-1}$, so $[G_2]=M^{-1} L^2 T^{-2}$, and one is unable to construct the gravitational radius from $G_2$, $c$ and mass of a point source $M$, so $g_{\mu\nu}$ can not depend on $r$ as there is no way to make it  dimensionless without a gravitational radius.
}
since in $(2+1)$ dimensions $G_{\mu\nu}=0$ implies $R_{\mu\nu\alpha\beta}=0$. In the same year Staruszkiewicz showed \cite{Staruszkiewicz} that the $(2+1)D$ spacetime with point masses is a Minkowski spacetime with nontrivial global properties. This fact was later rediscovered many times, see \cite{Gott1986} and references therein. Finally, in 1992 Banados, Teitelboim and Zanelli introduced \cite{BTZ} the concept of (2+1)-dimensional black hole (BTZ black hole) and started to investigate its global properties. BTZ black holes have been a subject of great interest among gravitationists since then, as their properties are often very similar to higher dimensional ones but BTZ are much more simple to study.

Isometric embeddings of GR manifolds have been used for the investigation of their geometric structure since the appearance of GR itself. Indeed, in 1921 Kasner found \cite{kasner3} the first embedding of Schwarzschild black hole in a (2+4)-dimensional flat ambient space. Unfortunately, this embedding was not global, i.e. it was not smooth at $0<r<+\infty$ (here and hereafter we will use the term "global" in this sense) and cannot be extended under the horizon. In 1959 Fronsdal found \cite{frons} a global embedding of Schwarzschild black hole, which turned out to cover the maximal analytic extension of the Schwarzschild metric. Moreover, there is an evidence that Kruskal knew this embedding when he was constructing his famous coordinates, but did not publish it. Many other embeddings of static black holes have been constructed since then \cite{fujitani,davidson,statja27,statja29,statja30}, but only a few of them are both minimal (i.e. the number of ambient space dimensions is equal to  minimal possible for this metric, see below) and global  \cite{statja40}.

The main goal of this paper is the construction and analysis of the global minimal embeddings of static black holes with negative cosmological constant in the background. The globality of embeddings is crucial when the embedding function is considered as a physical variable. A remarkable example of this approach is the description of gravity as the Regge-Teitelboim embedding theory \cite{regge,deser,davkar,statja18,statja25,faddeev,statja51}, where dynamics of gravity is formulated in terms of dynamics of curved surface isometrically embedded in flat ambient spacetime. In this theory the construction of explicit global embeddings of physically relevant metrics is required for the analysis of their possible non-Einsteinian modification \cite{statja26,statja33}. Another one of the main recent applications of explicit embeddings is the analysis of thermodynamic properties of manifolds with horizons.
This approach was proposed by Deser \& Levin \cite{deserlev99}, who discover a mapping between Hawking radiation detected in the spacetime with horizon and Unruh radiation encountered by observer who moves along the embedding surface of this spacetime in a flat ambient space.
The approach has been applied to many spacetimes with horizons (see, e.g., references in \cite{statja34}), for which many explicit embeddings have been constructed.

This paper is organized as follows. In Section \ref{secII} we briefly review main properties of the embedding procedure and the method of construction of exact embeddings proposed in \cite{statja27}. In Section \ref{secIII} we construct new explicit global embedding of static BTZ black hole in 6D spacetime.
In section \ref{secIV} we finish the proof of the absence of global 6-dimensional embeddings for AdS black holes which was started in \cite{statja40}. Then we construct  global 7-dimensional embedding for Schwarzschild-AdS black hole, as well as its generalization: the global embedding of $d$-dimensional Schwarzschild-AdS black hole in ($d+3$)D flat spacetime.

\section{The embedding of static black holes}\label{secII}

The basis of the embedding procedure is Janet-Cartan-Friedman theorem\cite{fridman61}:

\textit{An arbitrary n-dimensional pseudo-Riemannian manifold can be locally isometrically\linebreak
	embedded in $N$-dimensional pseudo-Riemannian
	spacetime of proper signature, if\linebreak \hbox{$N\geq n(n+1)/2$}.}

In the construction of exact embeddings of black holes the ambient space is assumed to be flat, which is consistent with the Regge-Teitelboim approach to gravity  where flat ambient space was preferred due to quantization reasons. However, embeddings in which ambient space is conformally flat can also be of some use, see \cite{Dunajski}.

For specific manifolds (e.g. with symmetry) the minimal required value of $N$ can be smaller than $n(n+1)/2$. For example, 4D spherically symmetric spacetimes can be locally embedded in 6D Minkowski space since their metric consists of two $2\times2$ blocks each of which can be embedded in 3D spacetime. The number $p=N_{\min}-n$ is called an embedding class and can be used as the invariant characteristic of the manifold.

Metric of the embedded spacetime becomes induced and can be expressed through the new geometric characteristic: an embedding function $y^a(x^\mu)$ which is a vector w.r.t. Poincare group of ambient space such that
\begin{align}\label{metric}
g_{\mu\nu}(x)=\partial_\mu y^a (x) \partial_\nu y^b (x) \eta_{ab},
\end{align}
where $x^\mu$ are coordinates on the manifold and $\eta_{ab}$ is a metric of ambient space. For example, the well-known embedding of $2$-sphere with metric interval $ds^2=r^2 (d\theta^2+\sin^2\theta d\varphi^2)$ looks like

\begin{align}\label{sphere}
& y^1 =r \cos \theta,\nonumber\\
& y^2 =r \sin \theta \cos \varphi,\\
& y^3 =r \sin \theta \sin \varphi. \nonumber
\end{align}

The system \eqref{metric} needs to be solved to construct an embedding. Unfortunately, in general case this system is very hard to solve explicitly, but in case of the metric with symmetries it is quite easier. Explicit embeddings of symmetric  manifolds can be constructed using a method proposed in \cite{statja27}.
Essentially, the idea of the method is to construct a surface,	which transforms into itself under the action
	of some subgroup of Poincare group of ambient spacetime,
	which is isomorphic
	(maybe locally)
	to the symmetry group of the metric.
	The transformations of the ambient spacetime and the symmetry of the metric
	could be the same, as in the above example: $SO(3)$ coordinate transformation corresponds to $SO(3)$ transformation of the embedding function, but it could be different as well. Simplest example of this difference is 2-dimensional embedding of
	the metric with translational symmetry:
	$y^1=\cos t, y^2=\sin t$. Here translations
	of the coordinate $t$ locally correspond to $SO(2)$ rotations of the plane $(y^1,y^2)$.
By listing all possible subgroups of ambient space Poincare group which are isomorphic to the symmetry group of metric one can obtain all possible types of surfaces with such metric. This procedure reduces the number of variables in the system \eqref{metric}, and if it becomes a system of ODEs, the problem often could be succesfully solved.

In case of spherically symmetric spacetimes this situation occurs when static \footnote{For spacetimes with a horizon it would be more correct to call them "invariant w.r.t. shifts of the coordinate $t$" since under the horizon $t$ is not timelike and the spacetime is not static in the strict sense. We will nevertheless call them static for the sake of brevity.} spacetimes are considered. In this case the problem is reduced to the embedding of $(t,r)$-block of the metric as the embedding of $(\theta,\phi)$-block can always be chosen as in \eqref{sphere}. Note that while this reduced anzats gives the exact answer only for static spherically symmetric spacetimes, the embedding of subspace of some non-symmetric spacetime is often interesting by itself. 
Embedding of such subspaces are called \textit{embedding diagrams} \cite{gr-qc/0305102}.

It was shown in \cite{statja27} that all possible symmetric (in the abovementioned sense) embeddings of such spacetimes fall into several types. These types are
\begin{align}
\begin{split}
y^0&=f(r)\sinh(\alpha t+u(r)),\label{vlo4}\\
y^1&= f(r)\cosh(\alpha t+u(r)),\\
y^2&=kt + h(r);
\end{split}
\end{align}
\begin{align}
\begin{split}
y^0&=f(r)\sin(\alpha t+u(r)),\label{vlo3}\\
y^1&=f(r)\cos(\alpha t+u(r)),\\
y^2&=k t+h(r);
\end{split}
\end{align}
\begin{align}
y^{0,1}=\frac{\alpha^2}{4}f(r)t^2+u(r)\pm f(r), \qquad y^2=\alpha f(r)t;\label{vlo6a}
\end{align}
and
\begin{align}
y^{0,1}=\frac{\alpha^2}{6} t^3+ \ls\al f(r)\pm \frac{1}{2}\rs t+u(r), \qquad y^2=\frac{\alpha}{2} t^2+f(r),\label{vlo6}
\end{align}
where the sign in \eqref{vlo6a} and \eqref{vlo6} is "$+$" for $y^0$ and "$-$" for $y^1$; $k$ and $\alpha$ are constant parameters and $h(r)$, $u(r)$ and $f(r)$ are functions which has to be determined by solving the system \eqref{metric}.
Embedding type \eqref{vlo4} with $k=0$ is called hyperbolic (the abovementioned Fronsdal embedding belongs to this type), the type \eqref{vlo3} --- spiral (or elliptic, if $k=0$; example of this type is the Kasner embedding), {the type \eqref{vlo6a} --- parabolic and type \eqref{vlo6} is called cusp (or cubic)}
embedding.

The construction and classification of embeddings for RN \cite{statja30}, Schwarzschild-de Sitter \cite{statja32} and RN-dS \cite{statja40} black holes was already done in our previous papers. The only remaining metric which has to be globally embedded in an ambient space with minimal possible number of dimensions is an embedding of the black hole (with or without an electric charge) with negative cosmological constant, i.e. Schwarzschild-AdS or RN-AdS one.

Unfortunately, it turned out that  it is impossible to embed these metrics in 6-dimensional flat ambient space. For S-AdS we had proven it earlier \cite{statja40}. RN-AdS case is more complicated one, so it took us a little time to realize that global embedding of RN-AdS black hole in 6-dimensional Minkowski space also does not exist.
{We will show it in Section~\ref{secIV-1}.
	As a result,}
we need to consider at least 7-dimensional ambient spacetime to proceed. It is worth noting that recently a 7-dimensional embedding of RN-AdS metric has been found \cite{soontae2018}, but it covers only the $r>r_-$ region of the manifold, so it is not global in our sense.

The main obstacle on the way of constructing {7-dimensional} embedding is the significantly increased number of symmetric surface types. As we said earlier, three of the seven components of the embedding function can be fixed uniquely due to $SO(3)$ symmetry, they correspond to the embedding of spherical part of the metric and coincide with \eqref{sphere}, whereas the dependence of other four on $t$ has to be determined by the abovementioned method.
In this case the number of realizations of time translation symmetry drastically increases in comparizon with \eqref{vlo4}-\eqref{vlo6} (for instance, one can consider all possible combinations of trigonometric and hyperbolic blocks in embedding function), and in this case the complete classification of surface types is much more tedious, so in this paper we do not want to perform it. Instead, we concentrate on the type of embedding which we found the most promising. To determine what type is it, let us turn to the 2+1-dimensional analog of Schwarzschild-AdS black hole, namely the static BTZ metric.

\section{Global embedding of static BTZ spacetime}\label{secIII}
Let us start with the general form of BTZ metric:
\begin{align}\label{BTZ_}
ds^2 =  \frac{(r^2 - r_+^2)(r^2 - r_-^2)}{l^2 r^2}dt^2 - \frac{l^2 r^2 dr^2}{(r^2 - r_+^2)(r^2 - r_-^2)} - r^2 \left(d\varphi - \frac{r_+ r_-}{l r^2} dt \right)^2,
\end{align}
where $r_+$ and $r_-$ are the horizons and $l$ is an AdS radius. The mass and the angular momentum are the following functions of the horizons:
\begin{align}
M = \frac{r_+^2 + r_-^2}{l^2}, \quad J = \frac{2r_+ r_-}{l}.
\end{align}

At first glance, there is no need to search for the embedding of \eqref{BTZ_} since shortly after the discovery of BTZ spacetime itself the following embedding\footnote{This is an embedding of a certain region of BTZ. Other parts are covered by similar embeddings; the junction between them is smooth. This behavior is typical for all hyperbolic embeddings; Fronsdal embedding also consists of two regions with smooth junction at the horizon.} was found \cite{BTZ2}:
\begin{align}\label{BTZ4}
\begin{split}
y^0 = l\sqrt{A(r)} \cosh	\hat{\varphi},\\
y^1 = l\sqrt{A(r)} \sinh	\hat{\varphi},\\
y^2 = l\sqrt{B(r)} \cosh	\hat{t},\\
y^3 = l\sqrt{B(r)} \sinh	\hat{t},
\end{split}
\end{align}
where $A(r)=\dfrac{r^2-r_-^2}{r_+^2-r_-^2}$, $B(r)=\dfrac{r^2-r_-^2}{r_+^2-r_-^2}$, $\hat{\varphi}=\dfrac{r_+\varphi-r_- t/l}{l}$, $\hat{t}=\dfrac{r_+t/l- r_-\varphi}{l}$.
\vskip 0.2em
Unfortunately, it is not periodic w.r.t. {angle} $\varphi$, so one must manually perform the identification $\varphi=\varphi+2\pi$ for it to become an embedding of BTZ spacetime (in fact, without such identification it is merely an embedding of AdS spacetime \cite{BTZ2}). To obtain this periodicity, one could replace hyperbolic functions in $y^0$ and $y^1$ with trigonometric functions of the argument $\varphi$. However, this transformation breaks the structure of the embedding \eqref{BTZ4} (see the footnote), where different regions of the manifold are described by different embeddings which are connected by smooth junction at the horizons. Therefore we consider two trigonometric blocks instead of one hyperbolic block, making the embedding 6-dimensional:
\begin{align}\label{BTZemb}
\begin{split}
& y^0=a^{-1}\sqrt{A(r)}\cos(at+\psi(r)),\\
& y^1=a^{-1}\sqrt{A(r)}\sin(at+\psi(r)),\\
& y^2=b^{-1}\sqrt{B(r)}\cos(bt+\chi(r)),\\
& y^3=b^{-1}\sqrt{B(r)}\sin(bt+\chi(r)),\\
& y^4 = r \cos \varphi,\\
& y^5 = r \sin \varphi.
\end{split}
\end{align}
With this ansatz we did not succeed in the construction of embedding for the generic BTZ metric since this anzatz does not allow a nonzero $g_{t\phi}$. However, we still able to construct a global embedding for the static case of BTZ (i.e. the analog of Schwarzschild-AdS black hole in 2+1 dimensions) with $J=0$ and $r_-=0$, for which the metric \eqref{BTZ_} reduces to
\begin{align}\label{BTZsp}
ds^2 = \left(-M+\frac{r^2}{l^2}\right)dt^2 - \dfrac{dr^2}{-M+\dfrac{r^2}{l^2}}  - r^2d\varphi^2.
\end{align}
In this case the condition $g_{tr}=0$ leads to
the following choice of function $\chi(r)$:
\begin{align}\label{chi}
\chi(r)=\frac{1}{2}\int \frac{A(r)b}{B(r)a} \ \psi'(r) dr.
\end{align}

Note that in {Schwarzschild-like} coordinates this metric {\eqref{BTZsp} is diagonal, but} has a
singularity at the horizon where $r=l\sqrt{M}$. To avoid this singularity we rewrite it in Eddington-Finkelstein coordinates:
\begin{align}\label{BTZsp2}
ds^2 = \left(-M+\frac{r^2}{l^2}\right)d\tilde t^2 - 2d\tilde t dr - r^2d\varphi^2.
\end{align}

The substitution of \eqref{BTZemb} and \eqref{BTZsp2} into \eqref{metric} gives a simple system of ODEs which can be satisfied by the following form of embedding function:
\begin{align}\label{BTZemb2}
\begin{split}
& y^0=\sqrt{1+\frac{1}{M}}\,\,r\cos\left(\frac{\tilde t}{l\sqrt{1+\dfrac{1}{M}}}\right),\\
& y^1=\sqrt{1+\frac{1}{M}}\,\,r\sin\left(\frac{\tilde t}{l\sqrt{1+\dfrac{1}{M}}}\right),\\
& y^2=\gamma{\sqrt{M}}{l}\cos\left(\frac{Mt+r}{\gamma M l}\right),\\
& y^3=\gamma{\sqrt{M}}{l}\sin\left(\frac{Mt+r}{\gamma M l}\right),\\
& y^4 = r \cos \varphi,\\
& y^5 = r \sin \varphi,
\end{split}
\end{align}
where the ambient space signature is $(++----)$ and $\gamma$ is an arbitrary positive constant. This embedding is smooth and real-valued at all
{$r>0$.} 

It must be noted that there are many types of embedding for the static BTZ metric \eqref{BTZsp}. For example, it is possible to embed it globally in the 2+3-dimensional ambient spacetime \cite{willison1011}, so our global embedding is not minimal. However, it allows the higher-dimensional generalization (in contrast with the 2+3-dimensional embedding), which is the subject of the section \ref{secIV-2}.

\section{Global embeddings of AdS black holes}\label{secIV}
\subsection{{The absence of 6-dimensional embeddings}}\label{secIV-1}
Let us turn to the consideration of the 3+1-dimensional black holes. As we mentioned earlier, the non-existence of 6-dimensional Schwarzschild-AdS black hole embedding was proven in \cite{statja40}, so we only need to analyse the remaining case of RN-AdS black hole which has the following metric:
\begin{align}
ds^2= F(r) dt^2-\dfrac{dr^2}{F(r)}-r^2 \ls d\te^2+\sin^2\te\, d\ff^2\rs,
\end{align}
where
\begin{align}\label{RNAdS}
F(r)=1-\dfrac{2m}{r}+\dfrac{q^2}{r^2}+\dfrac{r^2}{l^2}.
\end{align}

According to \cite{statja40}, the only possible types of global embedding for this metric are spiral \eqref{vlo3} and cusp \eqref{vlo6} ones.
The system \eqref{metric} gives the only condition to satisfy in order for cusp embedding to exist {(see \cite{statja40})}:
\begin{align}\label{ucusp}
u'^2=1-F+\dfrac{FF'^2}{4\alpha^2}.
\end{align}
After the substitution of \eqref{RNAdS} into \eqref{ucusp} some high order polynomial w.r.t. $r$ appears in the right-hand side. One might ask whether it is possible to find such $\alpha$ that this polynomial is non-negative for all $r$ and all $m$, $q$ and $l$ corresponding to non-extremal RN-AdS black hole.

It turns out that there are non-extremal black holes for which it is impossible. An example of such black hole is the one with  $q=2l$ and $m=4 l$.
The analysis of r.h.s. of \eqref{ucusp} in this case is given in Appendix~\ref{appB}. It shows that at any value of free parameter $\al$ this polynomial would not be non-negative at $r>0$. Taking \eqref{ucusp} into account, it means that cusp-type embedding of such metric is non-global.
Although there could be the other values of the black hole parameters for which r.h.s. of \eqref{ucusp} can be made positive, this lead us to a conclusion that there is no cusp embedding of RN-AdS black hole which is global for all possible values of physical parameters $m$, $q$ and $l$ corresponding to non-extremal black hole.

The non-globality of spiral embedding can be proved in similar way. In this case the system \eqref{metric} gives us the conditions {(see \cite{statja40})}:
\begin{align}\label{uza2}
4k^2(F+k^2)^2\ls u'-\frac{\al}{k}h'\rs^2=FF'^2-4\al^2(F+k^2)(F-1),
\end{align}
where $F(r)$ is defined as above and the parameter $k$ must be chosen in such way that $k^2\ge |F_{\min}|$. We chose the following values of physical parameters: $m=1,$ $q=1/2$ and $l=2$.
The analysis of r.h.s. of \eqref{uza2} in this case is also given in Appendix~\ref{appB}. It shows that this polynomial is not non-negative at any values of free parameters $\al$ and $k$.
Together with \eqref{uza2} it means
that the spiral embedding of non-extremal RN-AdS black hole which is global for all possible values of $m$, $q$ and $l$ does not exist.
This completes the classification of global 6-dimensional embeddings of static black holes.

\subsection{$d+3$ dimensional embedding of $d$-dimensional Schwarzschild-AdS black hole}\label{secIV-2}
Since global 6-dimensional embeddings for Schwarzschild-AdS and RN-AdS black holes do not exist, the minimal number of ambient space dimension has to be raised to 7. Unfortunately, the construction of 7-dimensional global embedding of RN-AdS black hole turns out to be quite cumbersome
due to the presence of two horizons.
For instance, the best known 7D embedding \cite{soontae2018} covers only $r_+$ horizon, but not both of them. Therefore we will restrict ourselves to the consideration of pure Schwarzschild-AdS case. The good news is that we eventually discovered that the construction of global embedding of $d$-dimensional Schwarzschild-AdS black hole in $d+3$-dimensional flat ambient spacetime can be solved for any $d\geq 4$, so we will immediately consider
{the metric of $d$-dimensional Schwarzschild-AdS black hole:}
\begin{align}\label{SadS}
ds^2=\left(1-\dfrac{R^n}{r^n}+\frac{r^2}{l^2}\right)dt^2-\dfrac{dr^2}{1-\dfrac{R^n}{r^n}+\dfrac{r^2}{l^2}}-r^2 d\Omega_{d-2}^2,
\end{align}
{
	where $R$ is the doubled mass of black hole,
	$d\Omega_{d-2}^2$ is a line element of $(d-2)$-sphere and $n=d-3$.
	Note that for Schwarzschild-AdS black hole at any values of the parameters has only one horizon.
}

Let us use the formulas \eqref{BTZemb} as a starting point in search for ($d+3$)-dimensional embedding of \eqref{SadS}. In this case the components $y^0,...,y^3$ are given by the same formulas:
\begin{align}\label{SAdSemb}
\begin{split}
& y^0=a^{-1}\sqrt{A(r)}\cos(at+\psi(r)),\\
& y^1=a^{-1}\sqrt{A(r)}\sin(at+\psi(r)),\\
& y^2=2b^{-1}\sqrt{B(r)}\cos(bt/2+\chi(r)),\\
& y^3=2b^{-1}\sqrt{B(r)}\sin(bt/2+\chi(r))
\end{split}
\end{align}
with signature $(++--)$ and $y^4,...,y^{d+2}$ are the embedding functions of $(d-2)$-sphere {analogous to \eqref{sphere}}.

One of the equations in the system \eqref{metric} is quite simple:
\begin{align}
g_{00}(r) = A(r)-B(r).
\end{align}
It is easy to see that if the functions $A$ and $B$ is chosen as follows:
\begin{align}\label{AB}
A(r)=1+\dfrac{r^2}{l^2},\qquad B(r)=\dfrac{R^n}{r^n}
\end{align}
that the expressions under the square roots in \eqref{SAdSemb} are strictly positive. The function $\chi(r)$ is still defined by \eqref{chi} and $\psi(r)$ has the form \begin{align}\label{psi}
\psi(r)=\int \dfrac{dr}{2bA(A-B)} \sqrt{(A-B) ( b^2 A'^2 B - 4 a^2 B'^2 A)+4a^2 b^2 (A+B+1)}.
\end{align} After the substitution of \eqref{AB} into \eqref{psi} it takes the form
\begin{align}
\psi(r) = \dfrac{R^{n/2}}{bl}\int dr\, \dfrac{r^{-1-3n/2}}{(r^2+l^2)}\dfrac{\sqrt{P(r)}}{g_{00}(r)}
\end{align}
where
\begin{multline}\label{P}
P(r) = r^2 b^2 (r R)^n l^4 +\\ + \ls l^2+r^2\rs\biggl( b^2 \ls 1-a^2 l^2\rs r^{2n+4} + l^2 \Bigl[ a^2 l^2 n^2 R^{2n}+\Bigl( b^2 r^2 \ls a^2 l^2 - 1\rs - a^2 n^2 \ls l^2+r^2\rs\Bigr) (r R)^n\Bigr]\biggr).
\end{multline}

The problem of construction of this embedding is essentially a problem of proving that  one can make the polynomial \eqref{P}  non-negative at $r>0$ by proper choice of the parameters $a$ and $b$. This proof is quite lengthy, so we put it in Appendix~\ref{appA} and write down the required form of $a$ and $b$:
\begin{align}
\begin{split}
a=\frac{1}{l}\sqrt{1-\frac{1}{\beta}},\qquad
&b=\frac{n}{l}\sqrt{\frac{\beta-1}{\al}},\\
\alpha>-1+\dfrac{(n+4)l^n}{R^n},\qquad
&\be>\left(2\alpha^2+\frac{1}{2}\right) \dfrac{R^n l^2}{r_0^{n+2}},
\end{split}
\end{align}
where we introduced two parameters $\al>0$ and $\be>1$. 	

Note that the integral in \eqref{psi} has a pole at $A=B$ (i.e. at the horizon
$r=r_0$). This singularity is a common feature of Schwarzschild coordinated and can be removed by the usual time transformation:
\begin{align}
t=\tilde t-
\dfrac{\sqrt{A(r_0)+B(r_0)+1}}{A(r_0)}
\int \dfrac{dr}{g_{00}(r)}.
\end{align}
It is easy to see that
when $\tilde t$ is used as the time coordinate
the embedding we constructed turns out to be smooth at $r>0$. Therefore it is global in our sense.

\section{Concluding remarks}
In this paper we constructed the new global embeddings for static BTZ metric and global minimal embedding for Schwarzschild-AdS metric. It may seem unsatisfying that we do not immediately succeed in the construction of embedding for a rotating BTZ spacetime, taking into account the relative simplicity of this metric and the fact that in this case the complete separation of variables in \eqref{metric} is possible. However, it should be noted that the desired globalness of the embedding drastically increases the difficulty of the problem. Since the rotating BTZ spacetime has two horizons, an additional increasing of ambient space dimension is probably required to smoothly cover both of them. Nevertheless, there is no evidence that this problem is unsolvable, and the construction of embedding for rotating $2+1$-dimensional black hole can probably give us some hints on the $3+1$-dimensional case, i.e. the Kerr metric. Note that the the only few embeddings for the Kerr metric are known; none of them is global and the number of ambient space dimensions in them is quite large: implicit Kuzeev embedding \cite{Kuzeev} is 9-dimensional and Hong \& Kim embedding \cite{gr-qc/0503079} is 14-dimensional.

Additional difficulties which are related to the existence of two horizons also arise when one tries to add an electric charge to the BTZ metric
or Schwarzschild-AdS metric.
For that reason the global embeddings of the charged BTZ and RN-AdS black holes require an additional study.

As we mentioned in the Introduction, global embeddings in Minkowski spacetime (which are often called GEMS) are extensively used in the analysis of the thermodynamical properties of manifolds with horizons. However, the complete globalness (i.e. smoothness at all $r>0$) of embeddings   usually is not required; it is necessary only in the neiborhood of the outer horizon.
It should be also noted that Hawking into Unruh mapping works only for embeddings of hyperbolic type \cite{statja34,statja36}. Since the new embeddings  of static BTZ and Schwarzschild-AdS spacetimes which were constructed here do not belong to this type (they are closer to the elliptic type), their usage would not lead to the appearance of this mapping. 

{\bf Acknowledgements.}
The work of D. S. and A. S. is supported by a grant from the Russian Foundation for Basic Research (Project No. 18-31-00169).

\begin{appendices}

\section{{Negative behavior of r.h.s. of \eqref{ucusp} and \eqref{uza2}}}\label{appB}
The r.h.s. of expression \eqref{ucusp} at the abovementioned values of physical parameters can be (up to {positive} constant multiplier) written as the following {polynomial}:
\begin{multline}\label{uza}
P(z,a) = z^{12}+(-a+1)z^{10}-4z^8+(8a+8)z^7+(-4a-56)z^6+\\+64z^5-160z^3+336z^2-256z+64,
\end{multline}
where $z=r/l$ and $a=\alpha^2$ is a free parameter.

Let us consider the values of this polynomial at the points $z_1=0.59$ and $z_2=3.5$. The approximate values of the appearing constants can be assumed safely:
\begin{equation}\label{sp1}
P(z_1,a) = -0.578+0.025 a,\qquad
P(z_2,a) = 3543670-231736 a.
\end{equation}
It is easy to see that the regions of possible values of $a$, in which these two quantities remain non-negative, do not overlap. It means that there is no such value of $a$ at which \eqref{uza} is non-negative.

The negativeness of r.h.s. of \eqref{uza2} can be proved
in a very similar way.
The r.h.s. of \eqref{uza2} at chosen values of physical parameters can be
written as the following polynomial (the positive constant multiplier is omitted as in above case):
\begin{multline}\label{sp2}
P(r,K,a)=(1-a)r^{12}+4(1-Ka-a)r^{10}+16ar^9+(-2a-1)r^8+\\+32(Ka+a+1)r^7+(-4Ka-68a-56)r^6+16(a+1)r^5+\\+(63-a)r^4-160r^3+84r^2-16r+1,
\end{multline}
where $a=4\alpha^2$, $K=k^2$ are free parameters.

{Let us consider the values of this polynomial at the points $r_1=0.2$ and $r_2=2$:}
\begin{align}
&P(r_1,K,a) = -0.01726-0.00042a+0.00015Ka,\label{sp3}\\
&P(r_2,K,a) = 8993-272a-256 Ka.\label{sp4}
\end{align}
The non-negativity of $P(r_2,K,a)$ implies $Ka\le 35.129-1.0625a$,
substitution of which into \eqref{sp3} leads to the condition $u(r_1,K,a)\le -0.01199 - 0.00058 a$,
which (as $a>0$) is incompatible with the assumption of non-negativity of $P(r_1,K,a)$.
Therefore the function \eqref{sp2} cannot be made non-negative by any possible choice of the free parameters $a$ and $K$.

\section{{Non-negativity of the polynomial \eqref{P}}}\label{appA}
Let us prove that for each integer $n\ge1$ the  parameters $a$ and $b$ can be chosen in such way that  \eqref{P} would be non-negative at all $r>0$.	Note that without loss of generality one can assume that $a,b\ge0$.	The polynomial \eqref{P} is positive at $r\to\infty$ if $a l <1$, so we assume it. Let us consider the quotient of $\eqref{P}$ and the positive quantity $b^2 (1-a^2 l^2) (l^2+r^2)$:
\begin{align}\label{P2}
\frac{P(r)}{b^2 (1-a^2 l^2) (l^2+r^2)}=h(r)+P_1(r),
\end{align}
where
\begin{gather}
P_1(r)=r^{2n+4}-\ls\alpha l^4+(\alpha+1)l^2 r^2\rs (r R)^n +R^{2n} l^4 \alpha,\\
h(r)=\dfrac{\be r^2 (r R)^n l^4}{l^2+r^2},\qquad  \alpha=\dfrac{\be a^2 n^2}{b^2}>0,\qquad 	\be=\dfrac{1}{1-a^2 l^2}>0.
\end{gather}
Note that the function $h(r)$ is strictly positive.

Let us examine $P_1(r)$. We introduce notations
\begin{align}
r=z^{1/n} l, \qquad R= x^{1/n} l
\end{align}
and consider the following function
\begin{align}
Z(z) = \frac{P_1(r)}{l^{2n+4}} = z^{2+4/n} -x(\alpha+1)z^{1+2/n}-x(z-x)\alpha,
\end{align}
instead of  $P_1(r)$.
Let us denote the only root of its second derivative as $z_2$, so $Z''(z_2)=0$. It can be found explicitly:
\begin{align}
z_2=\left(\dfrac{x(\alpha+1)}{n+4}\right)^{{n}/{(n+2)}}.
\end{align}
It is easy to see that since the first derivative $Z'(z)$ is positive at $z\to\infty$ and negative at $z=0$,	it also has only one root $z_1>z_2$, so the only extremum of the function $Z(z)$ is the minimum at $z_1>z_2$. For our purposes it is convenient to assume that $z_2>1$. To ensure it we put a lower boundary on $\alpha$:
\begin{align}
\alpha >\alpha_{min}= \begin{cases}
\dfrac{n+4}{x}-1, & x<n+4,\\
0, & x\geq n+4.	
\end{cases}
\end{align}

Consider the region $z>1$. Let us decompose $Z(z)$ in the following way:
\begin{align}
Z(z)=\left(z^{1+2/n}-x\ls\alpha+\frac{1}{2}\rs\right)^2 -x^2\ls\alpha^2+\frac{1}{4}\rs +x\alpha \left(z^{1+2/n}-z\right).
\end{align}
In this region the last term is positive, so the whole function is bounded from below. We also know that the minimum of $Z(z)$ is located here, so this boundary is valid for all $z>0$, and we have
	\begin{align}\label{spp1}
	P_1(r)>-l^{4}R^{2n}\ls\alpha^2+\frac{1}{4}\rs\equiv P_1^{\text{b}}.
	\end{align}

Consider the region $z<1$. Here we decompose $Z(z)$ other way:
\begin{align}
Z(z)=  z^{2+4/n} + x (\alpha+1)z(1-z^{2/n})  + \alpha x^2  -x (2\alpha+1)z.     \end{align}
In this region the first two terms is positive, so the whole function is positive if the last two terms is positive, i.e. if $z<z_0={x\alpha}/{(2\alpha+1)}$.
If $z_0>1$, the whole function is positive at $z<1$. We conclude that $P_1(r)$ is positive at $r<r_0$, where
\begin{align}
r_0=\text{min}\left(l, R\sqrt[n]{\dfrac{\alpha}{2\alpha+1}}\right).
\end{align}

Now let us consider the function $h(r)$, which is strictly positive and increases monotonically. We want to use
{the contribution of}
this function to compensate
a possible negative negative contribution from $P_1(r)$ in the region $r>r_0$.
Taking \eqref{spp1} into account, we need that
$h(r_0)+P_1^{\text{b}}>0$  to
ensure the positivity of \eqref{P2}. It is true if
\begin{align}
\be >    R^n(l^2+r_0^2) r_0^{-n-2}\ls\alpha^2+\frac{1}{4}\rs,
\end{align}
which is always possible. This
{condition}
can be slightly simplified by noting that $r_0\leq l$:
\begin{align}
\be > \left(2\alpha^2+\frac{1}{2}\right) \dfrac{R^n l^2}{r_0^{n+2}}.
\end{align}
We conclude that the non-negativity of $P(r)$ can be achieved if the following values of its free parameters $a$ and $b$ are chosen:
\begin{align}
\begin{split}
a=\frac{1}{l}\sqrt{1-\frac{1}{\beta}},\qquad
&b=\frac{n}{l}\sqrt{\frac{\beta-1}{\al}},\\
\alpha>-1+\dfrac{(n+4)l^n}{R^n},\qquad
&\be>\left(2\alpha^2+\frac{1}{2}\right) \dfrac{R^n l^2}{r_0^{n+2}},
\end{split}
\end{align}
where $\alpha$ and $\beta$ are new parameters of the embedding. Note that $a$ and $b$ are real, so $\al$ and $\be$ must be chosen properly, i.e.  $\al>0$ and $\be>1$. 	
\end{appendices}

\end{document}